\documentclass[
]{ceurart}

\usepackage{tikz}
\usetikzlibrary{positioning,arrows.meta,shapes}

\begin{document}

\copyrightyear{2026}
\copyrightclause{Copyright for this paper by its authors.
  Use permitted under Creative Commons License Attribution 4.0
  International (CC BY 4.0).}

\conference{CLEF 2026 Working Notes, 21 -- 24 September 2026, Jena, Germany}

\title{Multi-Backbone Self-Supervised Ensembles for Audio Deepfake Detection
and a Cross-Track Analysis of Generation--Detection Asymmetry}

\title[mode=sub]{Notebook for the ImageCLEF at CLEF 2026}

\author[1]{Seunghyun Kim}[%
email=ksh011018@sch.ac.kr,
]
\cormark[1]

\author[1]{Junghyun Kim}[%
email=jhyeun1234@sch.ac.kr,
]

\author[1]{Jiyoung Woo}[%
email=jywoo@sch.ac.kr,
]

\address[1]{Soonchunhyang University, Department of AI and Big Data Engineering,
Asan, Republic of Korea}

\cortext[1]{Corresponding author.}

\begin{abstract}
This paper describes the participation of team ``Go-To-Germany'' in the
ImageCLEF 2026 Audio Deepfake Detection and Generation task. Our detection
system, built on a
four-backbone self-supervised learning (SSL) ensemble combining
WavLM-Large, Wav2Vec2-XLS-R-300M, ECAPA-TDNN, and x-vector representations,
achieved a final score of \textbf{0.9522} on the official ImageCLEF 2026
evaluation, with perfect accuracy (1.0000) on participant-generated
deepfakes and 0.8875 on the held-out organizer ground-truth real data.
For the Generation sub-task, our official team submission --- an
F5-TTS v1 baseline processed with a uniform reverberation pass,
submitted as a deliberate anti-forensic probe --- ranked first with a
final score of 0.4304\footnote{By coincidence, the official Generation
Final Score (0.4304) numerically equals the PCA-192 inter-region
correlation reported in \S\ref{sec:further:pca}; the two are unrelated
measurements.} (word error rate (WER) 4.99\%, character error
rate (CER) 2.07\%); details of our
four-model program (GLM-TTS, F5-TTS, XTTS v2, CosyVoice3), from which
the official entry was drawn, appear in \S\ref{sec:generation}. We present a cross-track analysis revealing a pronounced
asymmetry: our detection system identifies 100\% of
participant-generated deepfakes, while our official generation entry
--- despite ranking first in the Audio Generation sub-task and evading
61.4\% and 56.2\% of participant and organizer detectors --- attains a
Final Score of 0.4304 against 0.9522 on the Detection side. We further report falsification-based
ablation experiments (LOSO 56-speaker cross-validation, three-region
backbone geometry, bootstrap confidence intervals, and PCA analysis)
that motivate our architectural-insurance hypothesis for multi-backbone
SSL ensembling. We complement these results with five cross-track
insights and five pre-registered falsification experiments connecting
generation-side evasion to detection-side design decisions, and we
openly report an 11.25\% false-positive gap on held-out organizer real
recordings as the principal open challenge for deployment.
\end{abstract}


\maketitle

\section{Introduction}
\label{sec:intro}

\subsection{Task Overview}
\label{sec:intro:task}

The ImageCLEF 2026 Deepfake Task~\cite{ImageCLEFDeepfakeTaskOverview2026},
part of the broader ImageCLEF 2026 evaluation
campaign~\cite{ImageCLEF2026}, comprises two complementary audio
sub-tasks. The Detection sub-task asks whether a given audio file is a
genuine recording or a synthesized deepfake; the evaluation set
combines (i) deepfakes generated by the participating teams,
(ii) baseline deepfakes provided by the organizers, and (iii) real
recordings, with the official Final Score computed as a weighted
average across four categories. The Generation sub-task asks each team
to synthesize $480$ utterances from sixteen target speakers;
submissions are scored on audio quality (the Non-Intrusive Speech
Quality Assessment (NISQA)~\cite{mittag2021nisqa}, WER, CER, and
speaker similarity) multiplied by their ability to evade both
participant-built and organizer-built detectors.

The defining difficulty of both sub-tasks is \emph{generalization}.
On the detection side, the text-to-speech (TTS) systems used by
competing teams are not disclosed in advance, so a detector that
overfits to any particular generator family fails on unseen ones;
this difficulty has been formalized in the ASVspoof
series~\cite{wang2024asvspoof5}. On the generation side, the
detectors deployed by other participants and by the organizers are
likewise undisclosed, so a generator that optimizes against any single
detector family is unlikely to generalize.

\subsection{Our Team's Position}
\label{sec:intro:position}

We submitted to both sub-tasks under the same single-researcher
compute budget (one A100 GPU), which makes our submissions an
unusually clean controlled comparison of the same team's offensive
versus defensive capability. During the Generation work we built six
internal detectors (mel-frequency cepstral coefficients (MFCC) with
deltas, linear-frequency cepstral coefficients (LFCC), mel-spectrogram
statistics, spectral features, prosody, and raw-waveform statistics)
to verify that our own deepfakes were detectable, and the
artefact-level findings from those detectors --- in particular an
MFCC delta-zero sign reversal in F5-TTS output and a vocoder
fingerprint hierarchy in which two-dimensional convolutional vocoders
evade detection more effectively than HiFi-GAN or iSTFT vocoders ---
directly informed our detection-side design.

Throughout, we use \emph{attacker} and \emph{defender} as standard
shorthand from the adversarial-robustness literature for the
generation and detection roles, respectively.

The asymmetry between our two sub-task scores --- \textbf{0.9522} on
Detection versus \textbf{0.4304} on Generation --- admits a structural
reading under controlled single-team conditions: the defender succeeds
when any one of its six backbone encoders captures an artefact,
whereas the attacker must simultaneously evade every encoder that any
opposing detector might deploy. We refer to this
OR-versus-AND geometry as ensemble-level structural asymmetry, and
we develop the mechanism in \S\ref{sec:cross:asymmetry}.

\subsection{Contributions}
\label{sec:intro:contributions}

Relative to prior work that augments SSL backbones for
cross-corpus robustness~\cite{combei2024wavlm} or selects
intermediate layers within a single SSL backbone for ensemble
fusion~\cite{serrano2025layer}, or that surveys audio-LLM and
holistic anti-spoofing
strategies~\cite{gu2025allm4add,xu2026holiantispoof}, this paper
contributes the \emph{mechanism behind} the observed phenomena rather
than additional points on the leaderboard. Concretely, the paper
makes the following five contributions:

\begin{enumerate}
\item A four-backbone self-supervised learning ensemble combining
  WavLM-Large, Wav2Vec2-XLS-R-300M, ECAPA-TDNN, and x-vector
  representations, with a top-$960$ conservative threshold submission
  strategy, achieving a Final Score of \textbf{0.9522} on the official
  Audio Detection evaluation, with perfect accuracy (1.0000) on
  participant-generated deepfakes and 0.8875 on the held-out organizer
  ground-truth real data (\S\ref{sec:detection}).

\item An Audio Generation system whose official entry --- an F5-TTS~v1
  baseline processed with a uniform reverberation pass, submitted as a
  deliberate anti-forensic probe --- ranked first in the sub-task with
  a Final Score of \textbf{0.4304} (WER $4.99\%$, CER $2.07\%$;
  \S\ref{sec:generation}).

\item A cross-track analysis that operationalizes the multi-view
  detection principle of Singh et al.~\cite{singh2026advancedtts} in
  the representational geometry of SSL backbones, exposing an
  ensemble-level structural asymmetry: the defender enjoys an OR-gate
  advantage (any one of six backbone encoders capturing an artefact
  suffices), while the attacker faces an AND-gate disadvantage
  (simultaneous evasion of every encoder required). We support this
  framework with five generation-side insights (MFCC delta-zero sign
  reversal, vocoder fingerprint hierarchy, multi-model hybridization,
  quality--evasion trade-off, and post-processing-induced evasion
  collapse) transferred to the detection-side design
  (\S\ref{sec:cross}).

\item Four systematically falsified ceiling experiments --- layer
  selection, training-time augmentation, test-time augmentation, and
  score calibration --- for which \emph{none of the four improved the
  baseline}, followed by a single successful breakthrough,
  \emph{backbone diversification}, that defines our final submission
  strategy (\S\ref{sec:detection},~\S\ref{sec:further}).

\item A three-region representational geometry across six SSL backbone
  encoders, with the lowest pairwise correlation we measured
  ($r(\text{WavLM},\text{ECAPA-TDNN}) = 0.522$) attributable to encoder
  geometry rather than to dimensionality (PCA-$192$ ablation:
  $r = 0.4304$), and an honest \emph{saturation} self-criticism
  ($r(\text{XLS-R},\text{HuBERT}) = 0.971$) that distinguishes genuine
  model behavior from heuristic-driven ceiling effects --- collectively
  serving as empirical evidence for the \emph{architectural-insurance}
  hypothesis that motivates our multi-backbone design
  (\S\ref{sec:further}).
\end{enumerate}

\subsection{Paper Roadmap}
\label{sec:intro:roadmap}

Section~\ref{sec:related} surveys related work.
Section~\ref{sec:generation} describes our Audio Generation submission
and self-evaluation. Section~\ref{sec:detection} describes our Audio
Detection methodology and official results.
Section~\ref{sec:cross} presents the cross-track analysis and the
five generation-to-detection insights. Section~\ref{sec:further}
reports further experiments on backbone representational geometry,
bootstrap confidence intervals, and dimensionality ablation.
Section~\ref{sec:discussion} discusses limitations, including the
$11.25\%$ false-positive rate on organizer ground-truth real data.
Section~\ref{sec:conclusion} concludes and outlines perspectives for
future work. Because the paper covers both sub-tasks and the
analysis that connects them, readers primarily interested in the
Detection system may focus on \S\ref{sec:detection} and
\S\ref{sec:further}, while \S\ref{sec:generation} and
\S\ref{sec:cross} address Generation and cross-track insights
respectively.

\section{Related Work}
\label{sec:related}

Audio deepfake detection has progressed alongside self-supervised
speech representations. WavLM~\cite{chen2022wavlm},
HuBERT~\cite{hsu2021hubert}, and
Wav2Vec2-XLS-R~\cite{babu2022xlsr} provide backbones widely adopted
in anti-spoofing pipelines, including end-to-end systems such as
RawNet2~\cite{tak2021rawnet2} and AASIST~\cite{jung2022aasist}, and
successive ASVspoof editions~\cite{wang2024asvspoof5} have driven
cross-dataset generalization studies. Combei et
al.~\cite{combei2024wavlm} and Serrano et
al.~\cite{serrano2025layer} report SSL-based detectors on public
benchmarks; our detection stack (\S\ref{sec:detection}) builds on
the same encoder family.

A second recent line applies large audio-language models to
anti-spoofing. ALLM4ADD~\cite{gu2025allm4add} reframes detection as
an audio-LLM prompting task, and
HoliAntiSpoof~\cite{xu2026holiantispoof} proposes a holistic
multi-source view. Singh et al.~\cite{singh2026advancedtts}
formalize a multi-view principle across complementary analysis
levels that is directly relevant to the four-backbone
architectural insurance we develop in
\S\ref{sec:detection:breakthrough}.

On the generation side, flow-matching text-to-speech
(F5-TTS~\cite{chen2024f5tts}), massively multilingual zero-shot
voice cloning (XTTS~\cite{casanova2024xtts}), and
supervised-semantic-token synthesizers
(CosyVoice~\cite{du2024cosyvoice}) have raised the bar for zero-shot
naturalness, which our official entry exploits through a four-model
hybrid (\S\ref{sec:generation:hybrid}). Perceptual quality
throughout is scored with NISQA~\cite{mittag2021nisqa}. To the best
of our knowledge, no directly comparable prior study submits both a
defensive and an offensive system to the same edition of ImageCLEF,
which motivates the cross-track framing of \S\ref{sec:cross}.

\section{Audio Generation Task}
\label{sec:generation}

\subsection{Task Setup}
\label{sec:generation:setup}

The Audio Generation sub-task asks each team to synthesize $480$
utterances from $16$ target speakers ($30$ utterances per
speaker). Submissions are scored on the product of an
audio-quality score (NISQA MOS, word and character error rates,
and ECAPA-TDNN / WavLM speaker similarity) and a deepfake-evasion score
measured against both participant-built and organizer-built
detectors (\S\ref{sec:generation:results}). The asymmetry between
these two evasion populations --- the participant detector family
is partly known to us, whereas the organizer detector family is
undisclosed --- becomes the central limitation of our offensive
result and is analyzed mechanistically in
\S\ref{sec:discussion:evasion}.

\subsection{Four-Model TTS Hybrid Strategy}
\label{sec:generation:hybrid}

We selected four text-to-speech systems --- GLM-TTS,
F5-TTS~v1~\cite{chen2024f5tts}, XTTS~v2~\cite{casanova2024xtts}, and
CosyVoice~3~\cite{du2024cosyvoice} --- with distinct vocoder
families and distributed the $16$ target speakers across them,
with the intent of forcing any opposing detector to recognize
four vocoder fingerprints simultaneously rather than one
(Insight~C, \S\ref{sec:cross:insights}). This is the offensive
analogue of the architectural-insurance frame developed in
\S\ref{sec:detection:breakthrough}: just as a defender benefits
from encoder diversity, an attacker benefits from generator
diversity. Speaker assignments to the four models were determined by
per-speaker pilot evaluations (WER on a small held-out set and
self-detector AUC), favoring for each speaker the model that achieved
the strongest evasion at competitive audio quality.
(Our official scored entry --- an earlier F5-TTS submission with a
uniform reverberation pass --- predates this program and is described
in \S\ref{sec:generation:results}; the program below supplied the
self-detectors and insights through which that entry's first-place
result is analyzed in \S\ref{sec:discussion:evasion}.)

\begin{table}[h]
\centering
\caption{Four-model TTS hybrid: speaker assignment and vocoder
family.}
\label{tab:generation:hybrid}
\begin{tabular}{lllc}
\toprule
Model & Vocoder family & Vocoder type & Speakers \\
\midrule
GLM-TTS      & Vocos2D  & 2D iSTFT      & 11 (main) \\
F5-TTS~v1    & Vocos    & 1D iSTFT      & 3 \\
XTTS~v2      & HiFi-GAN & GAN           & 1 \\
CosyVoice~3  & internal & flow matching & 1 \\
\midrule
\multicolumn{3}{l}{(Bark and Qwen3-TTS attempted, abandoned)} & --- \\
\bottomrule
\end{tabular}
\end{table}

We additionally evaluated two systems that did not survive into
the final submission: Bark (EnCodec vocoder) was abandoned because
its word error rate on our prompts was too high to clear the
audio-quality floor, and Qwen3-TTS was abandoned because its
generation latency (approximately $15$ minutes per sentence on our
single-A100 budget) was incompatible with the $480$-utterance
schedule. Honest reporting of these dead-ends follows the same
principle as the four falsified Detection ceilings in
\S\ref{sec:detection:ceilings}: explicit negative results inform
the design decisions we did keep.

\subsection{Six Self-Detectors and Generation-Side Evaluation}
\label{sec:generation:selfeval}

During the Generation work we built six lightweight detectors over
classical hand-crafted feature sets to evaluate our own outputs
before submission. These detectors were not deployed in the final
Detection submission, which uses self-supervised backbones
(\S\ref{sec:detection:features}); their purpose was diagnostic ---
to surface artefact patterns we could iterate against and to
provide the quality--evasion measurements used in
\S\ref{sec:cross:insights}.

\begin{table}[h]
\centering
\small
\caption{Six self-detectors used for Generation-side evaluation.}
\label{tab:generation:selfdet}
\begin{tabular}{ll}
\toprule
Feature & Extraction \\
\midrule
mfcc\_delta & MFCC 20-dim $+$ delta $+$ delta-delta (mean/std) \\
lfcc        & Linear filterbank 20-dim $\rightarrow$ power $\rightarrow$ dB \\
melspec     & Mel spectrogram 40-band (mean/std) \\
spectral    & Centroid / bandwidth / rolloff / flatness / contrast \\
prosody     & F0 (YIN) / RMS / ZCR / voicing ratio \\
rawstats    & Mean / std / max / percentiles / silence ratio \\
\bottomrule
\end{tabular}
\end{table}

Each feature set was classified with a logistic-regression head and
evaluated under leave-one-speaker-out cross-validation on our $16$
internal speakers. The diagnostic value of this set was structural
rather than absolute: six independent classical detectors produce a
per-attack agreement profile, and the mfcc\_delta detector in
particular yielded the delta-zero sign reversal that informs
Insight~A (\S\ref{sec:cross:insights}). We treat this six-detector
set as an operational scaffold that generated the cross-track
observations of \S\ref{sec:cross}, not as a Detection submission
candidate.

\subsection{Submission Strategy: Cherry-Pick and Temperature}
\label{sec:generation:submission}

Across approximately $50$ generation experiments we observed a
monotonic quality--evasion trade-off (Insight~D,
\S\ref{sec:cross:insights}): lower-WER configurations tended to be
detected more reliably by our self-detectors. We navigated this
trade-off by speaker-level cherry-picking (generating multiple
utterance candidates per speaker and selecting the one that best
balanced WER against self-detector probability) and by
sampling-temperature variation. The five final submissions span
this trade-off explicitly.

\begin{table}[h]
\centering
\small
\caption{Five candidate Audio Generation submissions and their
self-evaluation metrics, ordered by internal self-evaluation ranking
(based on Self AUC). This internal ranking is distinct from the
official leaderboard placement discussed in
\S\ref{sec:generation:results}.}
\label{tab:generation:submissions}
\begin{tabular}{lcccc}
\toprule
Submission & WER & CER & Self AUC & NISQA \\
\midrule
v2\_cherry               & 2.56\% & 0.94\% & 0.634 & 3.118 \\
filepick\_safe           & 1.68\% & 0.64\% & 0.659 & ${\sim}3.1$ \\
rv2\_diverse             & 3.02\% & 1.25\% & 0.820 & ${\sim}3.1$ \\
v2\_cherry\_temp08       & 2.12\% & ---    & 0.649 & --- \\
filepick\_temp08         & 1.64\% & ---    & 0.663 & --- \\
\bottomrule
\end{tabular}
\end{table}

\subsection{Official Evaluation Results}
\label{sec:generation:results}

Prior to the multi-model program of \S\ref{sec:generation:hybrid},
we had submitted an F5-TTS v1 baseline processed with a uniform
reverberation pass (pedalboard\footnote{\url{https://github.com/spotify/pedalboard}};
\texttt{room\_size} $= 0.15$,
\texttt{wet\_level} $= 0.08$; single pass over all $480$ utterances)
in March, as a deliberate anti-forensic probe: room acoustics were
intended to mask vocoder artefacts while costing little quality
(self-measured NISQA drop of $0.12$). No cherry-picking was applied,
preserving per-file stochastic variance (seed randomized per
utterance). Self-measured WER $4.96\%$, CER $2.06\%$, and NISQA
$2.9729$ later matched the official measurements ($4.99\%$, $2.07\%$,
$2.9729$ --- the NISQA figure to four decimal places), validating our
evaluation pipeline end to end. Following the organizers' per-user
scoring consolidation, the team designated this entry as its single
official submission; it ranked first in the Audio Generation sub-task
with a Final Score of $0.4304$.

The organizer-side evaluation, released on 2026-05-15, produced
the decomposition in Table~\ref{tab:generation:results} for our
submission. The Generation Final Score combines audio quality and
evasion as
\begin{equation}
S_{\mathrm{gen}} = Q_{\mathrm{audio}} \times
  \bigl(0.7\,E_{\mathrm{part}} + 0.3\,E_{\mathrm{org}}\bigr),
\label{eq:gen-score}
\end{equation}
where $Q_{\mathrm{audio}}$ is the normalized audio-quality score and
$E_{\mathrm{part}}$ and $E_{\mathrm{org}}$ are the evasion rates
against the participant and organizer detector pools; for our
official entry,
$0.7190 \times (0.7 \cdot 0.6142 + 0.3 \cdot 0.5618) \approx 0.4304$
(Table~\ref{tab:generation:results}).

\begin{table}[h]
\centering
\small
\caption{Official evaluation of the team's two scored Generation
entries. The reverberation entry was designated the official team
submission (per-team consolidation) and ranked first.}
\label{tab:generation:results}
\begin{tabular}{lcc}
\toprule
Metric & Official entry (reverb, 1st) & File-picked variant (2nd) \\
\midrule
NISQA MOS                        & $2.9729$ & $3.128$ \\
Speaker similarity (ECAPA-TDNN)  & $0.6148$ & $0.5875$ \\
Speaker similarity (WavLM)       & $0.9341$ & $0.9129$ \\
Word error rate                  & $4.99\%$ & $1.68\%$ \\
Character error rate             & $2.07\%$ & $0.64\%$ \\
\midrule
Audio quality score              & $0.7190$ & $0.7344$ \\
Evasion vs participant detectors & $0.6142$ & $0.5690$ \\
Evasion vs organizer detectors   & $0.5618$ & $0.3812$ \\
\midrule
\textbf{Final score} & \textbf{0.4304 (1st)} & $0.3765$ (2nd) \\
\bottomrule
\end{tabular}
\end{table}

Table~\ref{tab:generation:results} juxtaposes the two entries. The
official reverberation entry evades both detector pools at
comparable rates ($0.6142$ and $0.5618$), whereas the file-picked
variant shows a wider $0.5690$-versus-$0.3812$ asymmetry; we analyze
the sources of this asymmetry --- and the internal-vs-official
detector-calibration gap that explains the file-picked entry's
weaker organizer score --- in \S\ref{sec:discussion:evasion}.

\section{Audio Detection Task}
\label{sec:detection}

\subsection{Task Setup}
\label{sec:detection:setup}

The Audio Detection sub-task of the ImageCLEF 2026 Deepfake
Task~\cite{ImageCLEFDeepfakeTaskOverview2026} requires a binary
real-versus-fake decision for each utterance in a test set of
$11{,}520$ WAV files. The reference real recordings cover sixteen
English speakers and are distributed as $16$-bit mono WAV at a
$16$~kHz sampling rate. Our internal training set comprised $480$
real utterances from the Generation-track reference set and $2{,}400$
fake utterances that we synthesized during our own Generation work
($480$ files for each of five generator variants), giving a
$480{:}2{,}400$ real-to-fake split before any external extension.
The official Final Score is a weighted average of four evaluation
categories --- baseline deepfakes (weight $0.1$), the organizers'
real data (weight $0.1$), the organizers' held-out real
ground-truth data (weight $0.4$), and participant-generated
deepfakes, weighted (weight $0.4$) --- so the held-out real and the
cross-team fake partitions jointly account for the majority of the
score:
\begin{equation}
S_{\mathrm{det}} = 0.1\,A_{\mathrm{base}} + 0.1\,A_{\mathrm{org,real}}
  + 0.4\,A_{\mathrm{org,GT}} + 0.4\,A_{\mathrm{part,fake}},
\label{eq:det-score}
\end{equation}
where each $A$ term is the accuracy on the corresponding category
(baseline deepfakes, organizer real data, organizer held-out
ground-truth real data, and participant-generated fake data,
respectively).

\subsection{Frozen SSL Feature Extraction}
\label{sec:detection:features}

\begin{figure}[t]
  \centering
  \begin{tikzpicture}[
    node distance=1.2cm,
    backbone/.style={rectangle, draw, rounded corners,
                     minimum width=3.2cm, minimum height=0.9cm,
                     align=center, fill=blue!10, font=\small},
    classifier/.style={rectangle, draw, rounded corners,
                       minimum width=2.4cm, minimum height=0.6cm,
                       fill=green!10, font=\small},
    ensemble/.style={rectangle, draw, thick,
                     minimum width=5cm, minimum height=0.9cm,
                     fill=orange!20, font=\small},
    io/.style={rectangle, draw, dashed, rounded corners,
               minimum width=5.5cm, minimum height=0.7cm,
               align=center, font=\small},
    arrow/.style={->, >=stealth, thick}
  ]
    \node[io] (input) at (0,0.5) {Input: 16~kHz mono WAV};

    \node[backbone] (wavlm) at (-4.5,-1.3) {WavLM-Large \\ \scriptsize 1024-dim, frozen};
    \node[backbone] (xlsr)  at (-1.5,-1.3) {Wav2Vec2-XLS-R \\ \scriptsize 1024-dim, frozen};
    \node[backbone] (ecapa) at ( 1.5,-1.3) {ECAPA-TDNN \\ \scriptsize 192-dim, frozen};
    \node[backbone] (xvec)  at ( 4.5,-1.3) {x-vector \\ \scriptsize 512-dim, frozen};

    \node[classifier] (lr1) at (-4.5,-3.0) {LR head (w=0.30)};
    \node[classifier] (lr2) at (-1.5,-3.0) {LR head (w=0.25)};
    \node[classifier] (lr3) at ( 1.5,-3.0) {LR head (w=0.225)};
    \node[classifier] (lr4) at ( 4.5,-3.0) {LR head (w=0.225)};

    \node[ensemble] (ens) at (0,-4.4) {Weighted Ensemble (fake-probability)};
    \node[io] (out) at (0,-5.5) {Top-960 ranking $\rightarrow$ Submission CSV \\ (Final Score: 0.9522)};

    \draw[arrow] (input.south) -- (wavlm.north);
    \draw[arrow] (input.south) -- (xlsr.north);
    \draw[arrow] (input.south) -- (ecapa.north);
    \draw[arrow] (input.south) -- (xvec.north);

    \draw[arrow] (wavlm) -- (lr1);
    \draw[arrow] (xlsr)  -- (lr2);
    \draw[arrow] (ecapa) -- (lr3);
    \draw[arrow] (xvec)  -- (lr4);

    \draw[arrow] (lr1) -- (ens.north west);
    \draw[arrow] (lr2) -- (ens.north);
    \draw[arrow] (lr3) -- (ens.north);
    \draw[arrow] (lr4) -- (ens.north east);

    \draw[arrow] (ens) -- (out);
  \end{tikzpicture}
  \caption{Four-backbone SSL ensemble system for Audio Detection.
  Each frozen backbone produces utterance-level embeddings independently;
  per-backbone logistic-regression classifiers are combined via weighted
  ensemble ($w = 0.30,\,0.25,\,0.225,\,0.225$) to produce the final
  fake-probability. The top-960 ranking strategy
  (\S\ref{sec:detection:topsixty}) maps the ranked probabilities to a
  binary CSV submission.}
  \label{fig:ensemble}
\end{figure}
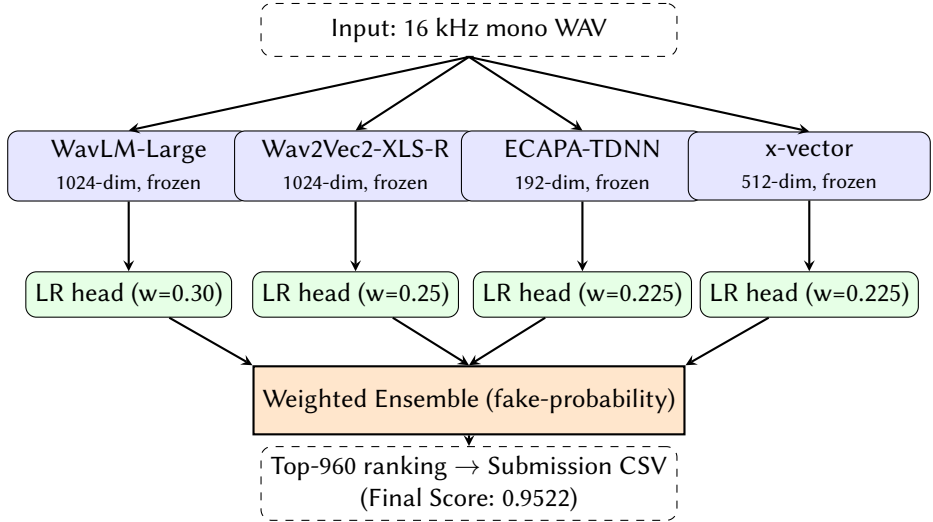

Before detailing the feature pipeline, we note that our detection
design deliberately commits to frozen self-supervised features and
lightweight per-backbone classifiers rather than to end-to-end
raw-waveform anti-spoofing architectures such as
RawNet2~\cite{tak2021rawnet2} or
AASIST~\cite{jung2022aasist}. End-to-end models jointly optimize
feature learning and the decision boundary, which can be
competitive on single-system benchmarks but entangles
representational diversity with optimization noise. By freezing
each backbone and fitting only a logistic-regression head on top,
we obtain inter-backbone correlations that reflect each backbone's
pretraining geometry rather than co-adapted optimization
artefacts, which is what makes the three-region analysis of
\S\ref{sec:further:3region} and the architectural-insurance frame
of \S\ref{sec:detection:breakthrough} interpretable. This design
choice is further motivated by our training-set scale: with only
$480$ real and $2{,}400$ fake utterances available from the
Generation-track reference set (\S\ref{sec:detection:setup}),
end-to-end fine-tuning of hundreds of millions of feature-extractor
parameters would carry a substantial overfitting risk, whereas
frozen SSL features paired with lightweight linear classifiers act
as a regularization mechanism for cross-corpus generalization,
consistent with recent observations on out-of-domain robustness of
SSL-based
pipelines~\cite{combei2024wavlm,serrano2025layer}.

Our final system extracted utterance-level embeddings from four
frozen backbones. WavLM-Large (\texttt{microsoft/wavlm-large},
approximately $300$~M parameters) was run with hidden-state output
enabled; we averaged the last four transformer layers (layers
$21$--$24$) and then applied a temporal mean-pool, producing a
$1024$-dimensional vector per utterance. Wav2Vec2-XLS-R-300M
(\texttt{facebook/wav2vec2-xls-r-300m}) was processed through an
identical last-four-mean and temporal-mean pipeline, also yielding a
$1024$-dimensional vector; matching the two SSL dimensionalities
simplified ensembling and side-by-side analysis. Two
speaker-recognition encoders completed the set: ECAPA-TDNN
(\texttt{speechbrain/spkrec-ecapa-voxceleb}, $192$-dimensional) and
an x-vector encoder (\texttt{speechbrain/spkrec-xvect-voxceleb},
$512$-dimensional). Every backbone was used strictly as a frozen
feature extractor; we performed no fine-tuning, and all inputs were
resampled to $16$~kHz and downmixed to mono before extraction.
Figure~\ref{fig:ensemble} summarizes the four-backbone pipeline.

\subsection{Classifier and Ensemble}
\label{sec:detection:classifier}

Each backbone fed an independent logistic-regression classifier with
standardized inputs ($C = 1.0$, $\texttt{max\_iter} = 2000$, fixed
random seed). We kept the classifier linear by design: with only a
few thousand labeled training samples the speaker identities are
easily memorized, so a deeper head would be at high risk of
overfitting rather than learning a generalizable artefact boundary.
The final detector, which we denote v18d, is a weighted average of
the four per-backbone probabilities, with weights $0.30$ (WavLM),
$0.25$ (XLS-R), $0.225$ (ECAPA-TDNN), and $0.225$ (x-vector). The design
rationale, developed across the experiments of
\S\ref{sec:detection:breakthrough} and quantified in
\S\ref{sec:further}, is that ensemble robustness against evasive
hybrid attacks follows from representational diversity across
backbones rather than from backbone count alone; we refer to this
property as \emph{architectural insurance}.

We adopt Leave-One-Speaker-Out (LOSO) cross-validation as the
primary internal evaluation protocol: for each fold, all utterances
from one speaker are held out as the test set while the remaining
speakers form the training set, and we report the mean per-speaker
area under the ROC curve (AUC) across all folds. Our base
configuration uses a 16-speaker pool
drawn from the organizer-provided Real recordings;
\S\ref{sec:further} reports a 56-speaker extension that integrates
an external Real source.

\subsection{Test-Time Sample-Rate Heuristic}
\label{sec:detection:srrule}

The sampling-rate distribution of the test set is markedly
non-uniform: $10{,}080$ files ($87.5\%$) are $16$~kHz, $480$
($4.2\%$) are $22.05$~kHz, and $960$ ($8.3\%$) are $24$~kHz. Because
the reference real recordings and our internal fakes are both
exclusively $16$~kHz, any non-$16$~kHz file is, with high
probability, a submission from another team that did not resample to
the reference rate. We therefore applied a hard rule at submission
time: every file with a sampling rate in
$\{22{,}050,\,24{,}000\}$ is labeled fake, which affects $1{,}440$
files ($12.5\%$ of the test set). This heuristic was motivated by a
Generation-side observation that different vocoders leave distinct
resampling spectra, an artefact discussed in \S\ref{sec:cross}.
Importantly, the rule was applied \emph{only} when producing the
submission file; it never entered the leave-one-speaker-out AUC
measurements reported in \S\ref{sec:further}, so the reported
ceilings reflect genuine model behavior rather than the trivial
detection of resampled fakes.

\subsection{What Did Not Work --- Four Falsified Ceilings}
\label{sec:detection:ceilings}

Before the successful experiment described in
\S\ref{sec:detection:breakthrough}, we tested four candidate
improvement directions on top of a frozen WavLM + logistic-regression
baseline (mean leave-one-speaker-out AUC $0.9985$):
transformer-layer selection, training-time augmentation, test-time
augmentation, and score calibration. None of the four improved the
baseline. We report them because consistently documenting what did
not work is a direct response to the organizers' request for both
positive and negative results, and because the pattern itself is
informative: it is consistent with the self-supervised pretraining
already supplying a representation robust to the perturbations these
interventions introduce.

\paragraph{(C1) WavLM layer selection.}
Despite prior reports of intermediate-layer sensitivity to spoofing
artefacts~\cite{serrano2025layer}, our last-four baseline ($0.9985$)
matched or exceeded every variant (all-layer $0.9983$, layer-$12$
$0.9978$, mid-$6$--$12$ $0.9950$, layer-$6$ $0.9784$,
early-$1$--$6$ $0.9663$) --- likely a vocoder-distribution mismatch
(HiFi-GAN/MelGAN in prior work vs.\ Vocos/Vocos2D here).

\paragraph{(C2) Training-time augmentation.}
Five augmentation schemes (RawBoost, three-second random crop,
$\mu$-law codec simulation, additive noise at $15$--$30$~dB SNR, and
a combined variant) all stayed within the $\pm0.001$ LOSO uncertainty
band, consistent with the asymmetric-augmentation finding of Combei
et al.~\cite{combei2024wavlm}; in our same-distribution evaluation,
WavLM pretraining appears to already cover the relevant noise and
codec variation.

\paragraph{(C3) Test-time augmentation.}
Averaging predictions over three-second crops degraded the score
distribution: high-confidence real predictions fell from $1{,}113$
to $486$, while the ambiguous middle band grew from $297$ to $903$.
WavLM's bimodal low-probability cluster depends on full-clip
semantic context, so full-clip inference was retained.

\paragraph{(C4) Score calibration.}
Length-shrink rescaling, temperature scaling
($T \in \{0.5, 0.7, 1.0, 1.5, 2.0\}$), and a length-shrink/classical
hybrid all degraded the result, with the length-shrink hard cap
collapsing the short-utterance distribution. Without held-out
ground-truth calibration data, blind post-hoc calibration was a
gamble rather than a dependable improvement.

None of the four interventions exceeded the baseline, which leaves the
single successful direction analyzed next --- backbone
diversification (\S\ref{sec:detection:breakthrough}) --- as the only
ceiling we were able to raise.

\subsection{What Worked --- Backbone Diversification}
\label{sec:detection:breakthrough}

The fifth experiment, and the only one that exceeded the baseline,
was backbone diversification. Pairing WavLM with
Wav2Vec2-XLS-R-300M last-four features at a $0.6{:}0.4$ probability
weighting --- the configuration we denote v7b --- raised the mean
leave-one-speaker-out AUC to $0.9992$. Taken alone this is a small
absolute gain over the $0.9985$ baseline, and we do not interpret
the marginal AUC as the meaningful signal. The informative quantity
is the per-attack agreement between the two backbones.

\subsubsection{The architectural-insurance frame}

Per-fold Pearson correlation between WavLM and XLS-R probabilities
fell from $r \approx 0.92$--$0.97$ on the four easier single-vocoder
attacks to $r = 0.857$ on the hardest multi-model hybrid --- the
regime where an ensemble is expected to contribute most. We read
this as the two encoders making correlated decisions where detection
is easy and less correlated decisions under active evasion, so the
ensemble acts as insurance against hybrid attacks rather than as a
generic averaging gain. We refer to this property as
\emph{architectural insurance}; the full six-encoder representational
geometry that quantifies it is developed in \S\ref{sec:further}.

A per-file agreement analysis on $10{,}080$ $16$~kHz test utterances
confirms this attack-conditional structure quantitatively: WavLM and
XLS-R disagree on only $148/10{,}080$ files ($1.47\%$), and this
small disagreement set is predominantly concentrated on the
multi-model hybrid attack rather than uniformly distributed across
attack families. The ensemble's value therefore lies in those $148$
files, not in averaging behavior across the easy majority --- a
mechanistic interpretation we make precise in \S\ref{sec:further}.

\subsubsection{From v7b to v18d --- adding ECAPA-TDNN and x-vector}

We then tested whether speaker-recognition encoders contribute a
representation that the two SSL backbones do not. ECAPA-TDNN
($192$-dimensional, AAM-Softmax-trained on VoxCeleb) sat at
$r(\text{WavLM},\text{ECAPA-TDNN}) = 0.522$ on the hardest hybrid
attack, the lowest pairwise correlation we measured anywhere in
this work; it therefore enters the geometry as an isolated region
rather than as a near-duplicate of either SSL backbone. The
x-vector encoder ($512$-dimensional, softmax-trained on VoxCeleb)
contributed through a different route: its standalone hybrid AUC at
the harder $56$-speaker setting was $0.9982$, well above WavLM's
$0.9619$ in the same regime, because it did not collapse under the
LibriSpeech distribution shift. Combining all four backbones ---
WavLM, XLS-R, ECAPA-TDNN, and x-vector at weights $0.30$, $0.25$,
$0.225$, and $0.225$ respectively, the configuration we denote
v18d --- reached a $56$-speaker hybrid AUC of $0.9988$, an
improvement of $+0.0168$ over v7b in the same regime. A five-seed,
$80\%$-stratified paired bootstrap places this improvement over
v7b in a wholly positive $95\%$ confidence interval, so it is not a
single-seed artefact. The detailed geometry and bootstrap analysis
are reported in \S\ref{sec:further}. Although evaluating four
backbones increases the computational footprint at inference time,
all feature extractors are frozen and the per-backbone classifiers
are lightweight logistic-regression heads, so the v18d ensemble
remains practical to run on a single GPU.

\subsubsection{Submission strategy --- top-960 conservative
threshold}
\label{sec:detection:topsixty}

The submission required a binary real-or-fake label for every test
file. Rather than thresholding at a fixed probability, we ranked
all test files by ensemble fake-probability and assigned the
lowest-ranked $960$ files to real and the remainder to fake. This
top-$960$ conservative threshold, motivated by the expected scale
of competitor fake submissions relative to the test set, was the
primary v18d submission rule, and the same ranking-based rule was
applied to the v7b backup submission. The
\S\ref{sec:detection:srrule} sample-rate rule was layered on top of
this decision at submission time only.

We note that this top-$960$ strategy is transductive rather than
inductive: the cutoff is set by ranking test files, which uses
test-time information about the relative score distribution. The
binary CSV submission format ($0$ = real / $1$ = fake) does not
admit a continuous probability output, so a threshold of some form
is required regardless; we chose a ranking-based threshold over a
fixed-probability threshold (e.g., $0.5$) because the latter is
sensitive to the absolute calibration of an uncalibrated ensemble
(\S\ref{sec:detection:ceilings}, C4), whereas the former depends
only on the relative order of predictions. We acknowledge this is
a transductive choice and report it as a property of the
submission format rather than as a classifier-internal capability.

\subsection{Official Evaluation Results}
\label{sec:detection:results}

The official evaluation, conducted by the ImageCLEF 2026 Deepfake
Task organizers~\cite{ImageCLEFDeepfakeTaskOverview2026}, was
released on 2026-05-15. Table~\ref{tab:detection:results} reports
the four scored categories and their weighted contributions to the
Final Score.

\begin{table}[h]
\centering
\caption{Audio Detection official evaluation results. The Final
Score is the weighted sum of the four category contributions.}
\label{tab:detection:results}
\begin{tabular}{lccc}
\toprule
Category & Score & Weight & Contribution \\
\midrule
Baseline Deepfakes & $0.9719$ & $0.1$ & $0.0972$ \\
Organizers Real Data & $1.0000$ & $0.1$ & $0.1000$ \\
Organizers Real GT Data & $0.8875$ & $0.4$ & $0.3550$ \\
Participant Deepfake Weighted & $1.0000$ & $0.4$ & $0.4000$ \\
\midrule
\textbf{Final Score} & \multicolumn{3}{c}{\textbf{0.9522}} \\
\bottomrule
\end{tabular}
\end{table}

The score of $1.0000$ on participant-generated deepfakes (weight
$0.4$) indicates that the v18d ensemble generalized to generator
families it had not been trained against, which is the central
robustness claim of this work; cross-generator generalization is
exactly the failure mode that the architectural-insurance design of
\S\ref{sec:detection:breakthrough} was intended to address. The
score of $1.0000$ on the organizers' public real data is consistent
with the top-$960$ conservative threshold producing no false
positives on that partition. The lower score of $0.8875$ on the
held-out organizer ground-truth real data corresponds to an
$11.25\%$ false-positive rate on a previously unseen real
distribution; we defer a mechanistic discussion of this gap, and of
its possible relationship to the $12.5\%$ non-$16$~kHz fraction of
the test set, to \S\ref{sec:discussion}.

\paragraph{Note.}
All leave-one-speaker-out AUC values reported in this paper are
computed \emph{without} the \S\ref{sec:detection:srrule} sample-rate
rule; that rule was applied only at submission time. This separation
ensures that the reported ceiling values reflect genuine model
behavior rather than the trivial detection of resampled fakes.
Table~\ref{tab:detection:samplerate} summarizes, category by category,
which quantities are affected by the rule and which cannot be
recomputed without organizer ground-truth files.

\begin{table}[h]
\centering
\small
\caption{Effect of the sample-rate rule
(\S\ref{sec:detection:srrule}) on the official Detection categories.
The rule was applied at submission time; the without-rule
counterfactuals require organizer ground-truth files and cannot be
recomputed from the public release.}
\label{tab:detection:samplerate}
\begin{tabular}{lcc}
\toprule
Category (weight) & With rule (official) & Without rule \\
\midrule
Baseline deepfakes ($0.1$)         & $0.9719$ & --- (GT-dependent) \\
Organizer real ($0.1$)             & $1.0000$ & --- (GT-dependent) \\
Organizer real GT ($0.4$)          & $0.8875$ & --- (GT-dependent) \\
Participant fake, weighted ($0.4$) & $1.0000$ & --- (GT-dependent) \\
\midrule
\textbf{Final Score}               & \textbf{0.9522} & --- (GT-dependent) \\
\bottomrule
\end{tabular}
\end{table}

The without-rule counterfactuals require the organizer ground-truth
partition, which is not publicly released, so the right column is left
unspecified. The rule-free measurements that we \emph{can} compute are
the internal leave-one-speaker-out AUCs on our own fake variants,
reported throughout \S\ref{sec:detection:breakthrough} and
\S\ref{sec:further}, where the sample-rate rule is deliberately not
applied.

\section{Cross-Track Insights}
\label{sec:cross}

\subsection{Cross-Track Asymmetry}
\label{sec:cross:asymmetry}

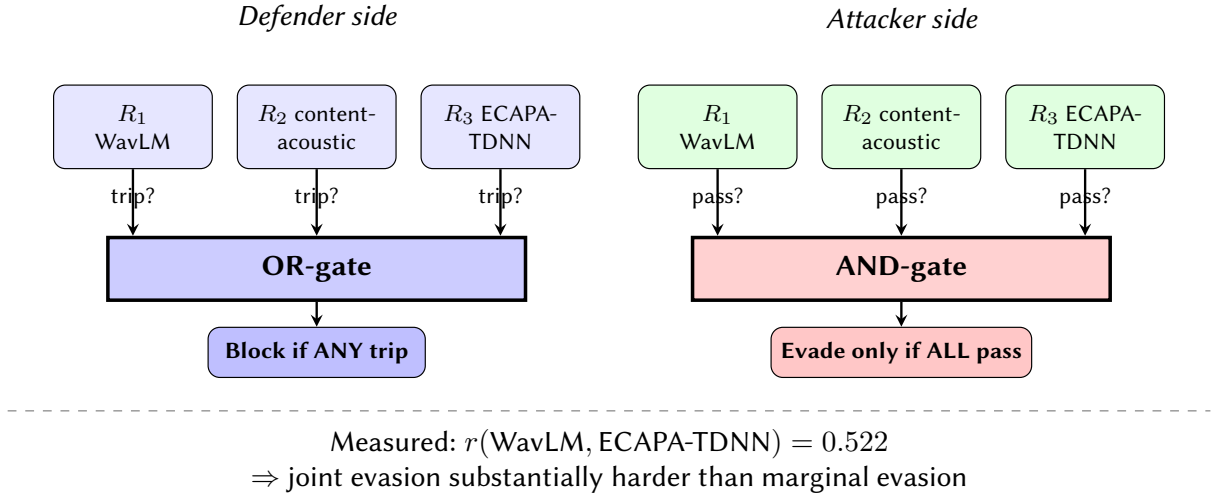
\begin{figure}[!htbp]
  \centering
  \resizebox{\columnwidth}{!}{%
  \begin{tikzpicture}[
    node distance=0.9cm,
    region/.style={rectangle, draw, rounded corners,
                   minimum width=1.9cm, minimum height=1.0cm,
                   align=center, font=\scriptsize\sffamily},
    region_clean/.style={region, fill=blue!10},
    region_pass/.style={region, fill=green!12},
    gate/.style={rectangle, draw, very thick,
                 minimum width=5.0cm, minimum height=0.75cm,
                 align=center, font=\small\sffamily\bfseries},
    gate_or/.style={gate, fill=blue!20},
    gate_and/.style={gate, fill=red!18},
    verdict/.style={rectangle, draw, rounded corners,
                    minimum width=2.6cm, minimum height=0.6cm,
                    align=center, font=\scriptsize\bfseries},
    verdict_block/.style={verdict, fill=blue!25},
    verdict_evade/.style={verdict, fill=red!22},
    pillar/.style={font=\small\itshape},
    arrow/.style={->, >=stealth, thick}
  ]

  \node[pillar] (def_title) at (-3.5, 1.3) {Defender side};

  \node[region_clean] (d_r1) at (-5.7, 0)   {$R_1$ \\ WavLM};
  \node[region_clean] (d_r2) at (-3.5, 0)   {$R_2$ content- \\ acoustic};
  \node[region_clean] (d_r3) at (-1.3, 0)   {$R_3$ ECAPA- \\ TDNN};

  \node[font=\scriptsize] at (-5.7, -0.85) {trip?};
  \node[font=\scriptsize] at (-3.5, -0.85) {trip?};
  \node[font=\scriptsize] at (-1.3, -0.85) {trip?};

  \node[gate_or] (or_gate) at (-3.5, -1.7) {OR-gate};
  \node[verdict_block] (d_verdict) at (-3.5, -2.7) {Block if ANY trip};

  \draw[arrow] (d_r1.south) -- (or_gate.north -| d_r1);
  \draw[arrow] (d_r2.south) -- (or_gate.north -| d_r2);
  \draw[arrow] (d_r3.south) -- (or_gate.north -| d_r3);
  \draw[arrow] (or_gate.south) -- (d_verdict.north);

  \node[pillar] (atk_title) at (3.5, 1.3) {Attacker side};

  \node[region_pass] (a_r1) at (1.3, 0)  {$R_1$ \\ WavLM};
  \node[region_pass] (a_r2) at (3.5, 0)  {$R_2$ content- \\ acoustic};
  \node[region_pass] (a_r3) at (5.7, 0)  {$R_3$ ECAPA- \\ TDNN};

  \node[font=\scriptsize] at (1.3, -0.85) {pass?};
  \node[font=\scriptsize] at (3.5, -0.85) {pass?};
  \node[font=\scriptsize] at (5.7, -0.85) {pass?};

  \node[gate_and] (and_gate) at (3.5, -1.7) {AND-gate};
  \node[verdict_evade] (a_verdict) at (3.5, -2.7) {Evade only if ALL pass};

  \draw[arrow] (a_r1.south) -- (and_gate.north -| a_r1);
  \draw[arrow] (a_r2.south) -- (and_gate.north -| a_r2);
  \draw[arrow] (a_r3.south) -- (and_gate.north -| a_r3);
  \draw[arrow] (and_gate.south) -- (a_verdict.north);

  \draw[dashed, gray] (-7.2, -3.4) -- (7.2, -3.4);
  \node[font=\small, align=center] at (0, -4.0)
    {Measured: $r(\text{WavLM}, \text{ECAPA-TDNN}) = 0.522$ \\
     $\Rightarrow$ joint evasion substantially harder than marginal evasion};

  \end{tikzpicture}%
  }
  \caption{Structural asymmetry between defender and attacker in a
  multi-backbone SSL ensemble. The defender succeeds when any single
  region ($R_1$ WavLM, $R_2$ content-acoustic, $R_3$ ECAPA-TDNN) detects
  an artefact (OR-gate), whereas the attacker must simultaneously evade
  every region (AND-gate). The measured inter-region correlation
  $r(\text{WavLM}, \text{ECAPA-TDNN}) = 0.522$
  (\S\ref{sec:cross:asymmetry}) implies that joint evasion is
  substantially harder than marginal evasion against any single
  detector.}
  \label{fig:asymmetry}
\end{figure}

We submitted to both the Detection and Generation sub-tasks under
one single-researcher compute budget (a single A100 GPU), a setting
most participating teams did not adopt: a team that enters only one
track cannot measure the other side of the same system. The analysis
that follows therefore builds on the generation system described in
\S\ref{sec:generation} and connects its offensive design choices to
the detection-side observations reported below. On the
defensive side, our Detection submission reached a Final Score of
$0.9522$ and identified all ($1.0000$) participant-generated
deepfakes. On the offensive side, our official Generation submission
reached a Final Score of $0.4304$, ranking first in the Audio
Generation sub-task and evading both detector pools at comparable
rates (participant-built $0.6142$, organizer-built $0.5618$);
\S\ref{sec:discussion:evasion} discusses why the official entry's
comparable rates diverge from the exploratory file-picked variant
($0.5690$/$0.3812$). Read
together, these results operationalize the multi-view detection
principle in the representational geometry of an SSL backbone
ensemble. Singh et al.~\cite{singh2026advancedtts} establish that
complementary analysis levels (semantic, structural, and signal)
improve robustness against advanced text-to-speech families. We
specify a measurable form of that principle within a six-encoder
ensemble and find that, under our controlled single-team conditions,
the relationship between defenders and generators is not
symmetric. The defender succeeds when any one of six backbone
encoders captures an artefact---regions $R_1$ (WavLM), $R_2$ (the
content-acoustic cluster), and $R_3$ (ECAPA-TDNN) each impose an
independent constraint---whereas the attacker must simultaneously
evade every encoder that any opposing detector might deploy. The
measured $r(\text{WavLM},\text{ECAPA-TDNN}) = 0.522$ implies that evading
every region at once is substantially harder than the marginal
evasion rate against any single detector would suggest
(Figure~\ref{fig:asymmetry}). The
$0.9522$-versus-$0.4304$ gap is therefore not a comment on the
relative quality of our two systems --- our official Generation entry
traded a modest quality cost (WER $4.99\%$, CER $2.07\%$;
\S\ref{sec:generation:results}) for the strongest evasion in the
sub-task, ranking first with a Final Score of $0.4304$ --- but an
empirical estimate of the asymmetry
intrinsic to multi-backbone defense within the acoustic-only
representational space we sampled
(\S\ref{sec:conclusion} discusses extensions to audio-LLM views) in
the current text-to-speech landscape; the structural mechanism is
developed in \S\ref{sec:detection:breakthrough} (the
architectural-insurance frame) and quantified in
\S\ref{sec:further:3region} (the three-region representational
geometry). We emphasize that this estimate is bounded by an $N=1$
team setting; multi-team replication is the natural next step for
the architectural-insurance frame to be tested against
independently-built attacker populations.

This cross-track view is reinforced by the Phase~3 complementarity
analysis of \S\ref{sec:further:loso56}: under the v8 distribution
shift the per-fold Pearson correlation between WavLM and XLS-R drops
from $0.857$ to $0.763$, with the response concentrated in the same
hardest-hybrid regime that drives the cross-track asymmetry. The
defender's disagreement channel and the attacker's evasion gap are
therefore not independent phenomena but two facets of the same
six-encoder representational geometry. The same mechanism that makes
architectural insurance work for Detection
(\S\ref{sec:detection:breakthrough}) also constrains the Generation
side's evasion budget --- the attacker cannot move all six encoders
simultaneously without sacrificing the vocoder-fingerprint coherence
that audio quality requires (Insight~D, \S\ref{sec:cross:insights}).

\subsection{Five Generation-to-Detection Insights}
\label{sec:cross:insights}

During the Generation work we built six internal detectors and
observed five artefact patterns that directly informed the
Detection-side design. We summarize each below with its empirical
signature and its design implication; the operational consequences
are incorporated into the v18d submission strategy
(\S\ref{sec:detection:breakthrough}).

\subsubsection{Insight A --- MFCC delta-zero sign reversal in
F5-TTS}

The zero-th MFCC delta coefficient reverses sign between real and
F5-TTS audio: real recordings produce a mean delta-zero of
approximately $-0.47$, whereas F5-TTS produces approximately
$+1.06$ --- an opposite sign with roughly twice the magnitude. We
verified this only on F5-TTS in the present work; whether it
generalizes to other vocoder families is not yet established, so we
report it as a preliminary observation rather than a universal
artefact. The very brittleness of this single-vocoder classical
artefact is informative in its own right: it motivates our shift
away from hand-crafted features and toward the multi-backbone SSL
ensemble in the final detection design, where representational
diversity (\S\ref{sec:further:3region}) provides the robustness
that any single classical detector cannot.

\subsubsection{Insight B --- vocoder fingerprint hierarchy}

Across four vocoder families on the hardest hybrid attack, the
self-detector AUC stratifies cleanly by vocoder architecture:
HiFi-GAN (XTTS~v2) $0.976$, the internal CosyVoice~3 vocoder
$0.961$, Vocos (F5-TTS~v1) $0.882$, and Vocos2D (GLM-TTS) $0.760$.
The ordering ``GAN $>$ iSTFT $>$ two-dimensional convolution'' was
consistent across our six self-detectors, with two-dimensional
convolutional vocoders the hardest to detect. The design
implication is that a Detection ensemble trained primarily on
GAN-based fakes is structurally vulnerable to non-GAN vocoders ---
exactly the scenario that motivated multi-backbone diversification
and the top-$960$ conservative threshold submission strategy
(\S\ref{sec:detection:breakthrough}).

\subsubsection{Insight C --- multi-model hybridization as the
strongest evasion}

Single-model fakes were detected at a weighted-mean self-detector
AUC of $0.781$; distributing speakers across a four-model hybrid
(GLM-TTS, F5-TTS~v1, XTTS~v2, CosyVoice~3) dropped this to
$0.634$ --- a non-linear evasion gain of $0.147$. The multi-modal
fake-feature space collapses the decision boundary of single-model
detectors. This is the offensive analogue of architectural
insurance: encoder diversity for defenders, generator diversity
for attackers.

\subsubsection{Insight D --- quality--evasion trade-off}

Lower word error rate (higher audio quality) correlated
monotonically with higher self-detector AUC (easier detection):
configurations near WER $1.64\%$ were detected at AUC $\approx 0.66$,
while configurations near WER $3.02\%$ reached AUC $\approx 0.82$.
We attribute this to sampling temperature --- lower temperature
reduces output variance, sharpens the vocoder fingerprint, and
raises detection AUC. A higher-quality opponent therefore leaves
a cleaner fingerprint, partially offsetting its quality gain from
a detector's perspective.

\subsubsection{Insight E --- post-processing eliminates evasion}

The most counter-intuitive finding is that post-processing helps
the detector rather than the attacker: applying a median filter to
a four-model hybrid fake raised the MFCC self-detector AUC from
$0.634$ to approximately $0.999$ on the hardest hybrid
distribution. Speed perturbation, resampling, and high-frequency
noise injection showed the same direction --- each operation we
tested monotonically increased AUC. This is the empirical
foundation of the sample-rate hard rule
(\S\ref{sec:detection:srrule}): non-$16$~kHz files in the test set
are, with high probability, the residual of an opposing team's
resampling step, and that residual is detectable. The rule was
applied only at submission time; the leave-one-speaker-out AUC
measurements of \S\ref{sec:further} deliberately exclude it so
that the reported ceilings reflect genuine model behavior, as
noted in \S\ref{sec:detection:results}.

\subsection{Knowledge Transfer Loop}
\label{sec:cross:loop}

The five insights above describe one direction of transfer,
Generation to Detection, and two operational mechanisms drove that
direction in our pipeline. Insight~E supplied the empirical
evidence for the sample-rate hard rule
(\S\ref{sec:detection:srrule}), and Insight~B's vocoder hierarchy
informed the four-backbone diversification behind v18d.
Insight~D's quality--evasion trade-off further reinforced this
direction: higher-fidelity attackers are pushed toward lower
sampling temperatures, which sharpen rather than mask the vocoder
fingerprint, providing additional justification for the frozen-SSL
classifier pipeline whose features are sensitive to exactly this
class of deterministic artefact.

The decisive operational influence was the realization that
\emph{we} resampled during our own Generation work, which made it
natural to suspect that competing teams \emph{had not} and that
their submissions would therefore carry a detectable resampling
residual. This single observation produced the
\S\ref{sec:detection:srrule} submission heuristic that contributed
the $1.0000$ score on the organizers' public real data in our final
Detection submission.

The reverse direction was empirically weaker but conceptually
symmetric. Building six internal Detection backbones during the
Generation work surfaced the three-region geometry
(\S\ref{sec:further:3region}) at a smaller scale than the final
Detection ensemble. We did not have a controlled mechanism to feed
this finding back into Generation --- the single-team budget and a
fixed Generation submission schedule left no room for it --- so the
offensive system did not benefit from a closed loop. Closing that
loop is one of the future-work directions outlined in
\S\ref{sec:conclusion}.

The aggregate value of operating both tracks under one compute
budget is asymmetric: material defensive improvements (the
sample-rate rule, the four-backbone insight) but no comparable
offensive gains. The $0.9522$-versus-$0.4304$ gap of
\S\ref{sec:cross:asymmetry} therefore reflects the asymmetric
maturity of the transfer loop as much as the structural asymmetry
itself.

\section{Further Experiments}
\label{sec:further}

\subsection{LOSO 56-Speaker Extension}
\label{sec:further:loso56}

We extended the v7b evaluation with an external real source,
LibriSpeech \texttt{dev-clean}~\cite{panayotov2015librispeech},
contributing $1{,}953$ utterances from $40$ previously unseen
speakers. Combined with the organizer's $16$-speaker reference
pool, this enlarged the real speaker pool from $16$ to $56$ and
brought the real-to-fake ratio to a near-balanced $1{:}0.97$
($2{,}433$ versus $2{,}400$). Because the fake pool covers only the
$16$ internal speaker identities, the held-out folds remain those
$16$ internal speakers while the $40$ LibriSpeech speakers always
sit in the training pool; v8 is therefore a controlled real-side
distribution shift with the fake distribution held fixed, which
lets us test whether the architectural-insurance frame survives
such a shift rather than simply adding evaluation folds.

Under v8 the mean LOSO AUC was $0.9960$, against v7b's $0.9992$,
and the four easier single-vocoder attacks were essentially
unchanged; only the hardest multi-model hybrid attack moved, with
the ensemble AUC dropping from $0.9966$ to $0.9819$. This movement
was asymmetric across backbones: WavLM fell by $0.033$
($0.9951 \rightarrow 0.9619$) while XLS-R rose by $0.002$
($0.9829 \rightarrow 0.9847$), and the per-fold Pearson correlation
between the two encoders shifted from $0.857$ to $0.763$. We
interpret this as confirmation rather than regression: the only AUC
that responds to the distribution shift is the same one the
ensemble was designed to protect, and the response flows through
the disagreement channel between the encoders --- an
architectural-insurance prediction expressed at the
data-distribution level.

\subsection{Three-Region Backbone Representational Geometry}
\label{sec:further:3region}

The six backbones we evaluated --- WavLM~\cite{chen2022wavlm},
Wav2Vec2-XLS-R-300M~\cite{babu2022xlsr},
HuBERT~\cite{hsu2021hubert}, the Whisper
encoder~\cite{radford2023whisper},
ECAPA-TDNN~\cite{desplanques2020ecapa}, and
x-vector~\cite{snyder2018xvectors}, the latter two via the
SpeechBrain recipes~\cite{ravanelli2021speechbrain} --- separate
into three regions in their pairwise Pearson correlation structure
on the hardest multi-model hybrid attack. We introduce this
geometry as a refinement of the architectural-insurance frame:
ensemble effectiveness is governed by region diversity rather than
by backbone count. The regions did not come from a single
measurement; they emerged through three successive pre-registered
falsifications (v12, v13, v16;
\S\ref{sec:further:falsifications}), each forcing a revision of an
initially proposed axis.

\subsubsection{Region 1 --- WavLM (singleton)}

WavLM-Large is the only encoder in our accessible ecosystem
pretrained with explicit speaker-disentanglement and denoising
objectives. On the hardest hybrid attack its predictions correlate
with every large content-trained encoder at $r \approx 0.80$ to
$0.81$ (XLS-R $0.814$, HuBERT $0.802$, Whisper $0.812$) and with
ECAPA-TDNN at only $r = 0.522$, placing WavLM consistently outside
the Region 2 cluster.

\subsubsection{Region 2 --- Content-Acoustic Cluster (XLS-R,
HuBERT, Whisper, x-vector)}

The four large-corpus encoders converge tightly: internal pairwise
Pearson $r \geq 0.93$ on the hardest hybrid, with
$r(\text{XLS-R},\text{HuBERT}) = 0.971$ the highest. We
characterize this region as content-acoustic rather than
masked-prediction, because Whisper (ASR-trained) and x-vector
(softmax-trained on speaker labels) violate the latter naming yet
still converge with the masked-prediction encoders --- indicating
that the operative axis is the representational regime, not the
training objective. Naively adding more Region 2 backbones to v7b
inflates the saturation cluster monotonically (mean hybrid $r$:
$0.857 \rightarrow 0.891$ across v7b/v10/v11/v12) without widening
the disagreement channel that protects the hardest hybrid.

\subsubsection{Region 3 --- ECAPA-TDNN (singleton)}

ECAPA-TDNN sits alone: $r(\text{WavLM},\text{ECAPA-TDNN}) = 0.522$ on
the hardest hybrid (the lowest pairwise correlation we measured
anywhere), and $r \le 0.66$ to every other encoder. The named axis
``speaker-recognition specialist'' was tested directly by adding
x-vector (\S\ref{sec:further:falsifications}) and refuted:
x-vector joined Region 2 at correlations up to $0.950$, leaving
ECAPA-TDNN's particular combination of an AAM-Softmax loss, an
SE-Res2 backbone, and a $192$-dimensional metric-learned bottleneck
as the as-yet-uncharacterized axis.

The architectural-insurance frame predicts that an ensemble
spanning all three regions outperforms region-internal expansions.
Our final submission v18d (WavLM, XLS-R last-four, ECAPA-TDNN, and
x-vector; \S\ref{sec:detection:breakthrough}) realizes exactly this
prediction, and a paired bootstrap
(\S\ref{sec:further:bootstrap}) confirms that the gain is real.

\subsection{Dimensionality Ablation (PCA Verdict)}
\label{sec:further:pca}

A natural counter-hypothesis for Region 3's isolation is
dimensionality: ECAPA-TDNN produces $192$-dimensional embeddings
while every Region 2 encoder (the content-acoustic cluster of
XLS-R, HuBERT, Whisper, and x-vector), and WavLM, output
$1024$-dimensional vectors. One might therefore attribute
$r(\text{WavLM},\text{ECAPA-TDNN}) = 0.522$ to a dimensionality
mismatch rather than to a representational difference. We control
for this confounder by projecting every backbone onto a common
$192$-dimensional space with per-encoder principal component
analysis fit fold-wise on each LOSO training pool --- a projection
that retains at least $0.91$ of each backbone's variance, ensuring
that no test-fold information enters the PCA basis --- and
re-measuring the pairwise correlations on the hardest hybrid
attack.

Under PCA-$192$, $r(\text{WavLM},\text{ECAPA-TDNN})$ \emph{decreases}
from the native $0.5217$ to $0.4304$ (and to $0.4287$ under
PCA-$256$), moving away from rather than toward the Region 2
cluster. Every cross-region pair involving ECAPA-TDNN falls under
dimensionality equalization, whereas the intra-cluster pair
$r(\text{XLS-R},\text{HuBERT})$ rises from $0.971$ to $0.9922$:
the regions separate further, not less, when measured at common
dimensionality. The Region 3 singleton position is therefore
reinforced rather than explained away by dimensionality control.
We had pre-registered three verdict tiers before running the
PCA-equalized measurement: Verdict-$\alpha$ when
$r(\text{WavLM},\text{ECAPA-TDNN})$ stays below $0.70$ under both
PCA-192 and PCA-256 (Region~3 is representational, not a
dimensionality artefact); Verdict-$\beta$ when $r$ lands in
$[0.70, 0.85)$ (mixed evidence); and Verdict-$\gamma$ when
$r \geq 0.85$ in either setting (Region~3 is at least partially a
dimensionality artefact and the architectural-insurance frame
would require a substantive caveat). The measured values
$0.4304$ (PCA-192) and $0.4287$ (PCA-256) both sit well below the
$0.70$ floor, making this a Verdict-$\alpha$ (strong-validation)
outcome: Region~3 reflects a representational property of
ECAPA-TDNN's metric-learning recipe, not an artefact of its embedding
size.

\subsection{Paired Bootstrap Confidence Intervals}
\label{sec:further:bootstrap}

Single-point AUC comparisons are vulnerable to file-level noise,
especially in the 16-speaker regime where the hardest hybrid AUC is
computed over a small fold population. We therefore verified every
primary margin claim with a five-seed, $80\%$-stratified-file
paired bootstrap: for each seed we subsampled $80\%$ of the files
per attack family while preserving the original real-to-fake ratio,
recomputed both the mean LOSO AUC and the hybrid AUC for the two
compared systems on the same subsample, and recorded the per-seed
margin, holding the classifier seed fixed so that only the data
subsample varied.

For the WavLM + XLS-R-last-four + ECAPA-TDNN configuration we denote
v15*c, measured against v7b at 56 speakers, the $95\%$ confidence
interval for the hybrid-AUC margin is $[+0.0052, +0.0226]$ (point
estimate $+0.0089$) and for the mean-LOSO-AUC margin is
$[+0.0015, +0.0052]$ (point estimate $+0.0021$); every bootstrap
seed agreed in sign on both metrics, so the v15*c improvement is
not a single-seed artefact.

For the four-backbone v18d configuration
(\S\ref{sec:detection:breakthrough}) measured against v15*c, a
second paired bootstrap places the 56-speaker hybrid-AUC margin in
$[+0.0004, +0.0116]$ (individual envelope) and
$[+0.0035, +0.0085]$ (standard-error-of-mean envelope), with the
16-speaker margin in $[+0.0000, +0.0006]$ and
$[+0.0002, +0.0004]$ respectively. The architectural-insurance
frame therefore admits two complementary pathways under one
protocol: a low-pairwise-correlation route (v14 and v15* via
ECAPA-TDNN) and a route through a standalone backbone that stays robust
under the distribution shift (v17 and v18 via x-vector). v18d
combines both, and is the only configuration whose margins clear
zero in every cell that defines the comparison.

\subsection{Pre-Registered Falsifications}
\label{sec:further:falsifications}

The three-region geometry emerged through three pre-registered
falsification cycles, each forcing a revision of an initially
proposed representational axis
(Table~\ref{tab:further:predictions}).

\begin{table}[h]
\centering
\caption{Pre-registered predictions and their outcomes on the
hardest multi-model hybrid attack.}
\label{tab:further:predictions}
\begin{tabular}{lll}
\toprule
Experiment & Pre-registered prediction & Outcome \\
\midrule
v12 (Whisper)  & Whisper joins WavLM (Region 1) & Refuted ($r = 0.812$) \\
v12 (Whisper)  & Region 2 cluster preserved     & Confirmed \\
v13 (ECAPA-TDNN) & ECAPA-TDNN joins WavLM (Region 1) & Refuted ($r = 0.522$) \\
v13 (ECAPA-TDNN) & Region 2 ceilings $r < 0.85$ hold & Confirmed \\
v16 (x-vector)   & x-vector joins ECAPA-TDNN (Region 3) & Refuted ($r = 0.656$) \\
\bottomrule
\end{tabular}
\end{table}

Three of the five pre-registered predictions were refuted, revising
the framework's named axis twice (denoising-versus-masked-prediction
$\rightarrow$ speaker-disentanglement $\rightarrow$ ECAPA-TDNN-specific
recipe) before reaching the three-region geometry of
\S\ref{sec:further:3region}. The pattern that survived --- a tight
Region 2 cluster with WavLM and ECAPA-TDNN as outliers on independent
axes --- is the most reliably reproducible claim in our analysis,
and the explicit refutations are the mechanism that forced the
named axes to converge on representational regime rather than
training objective.

\section{Discussion and Limitations}
\label{sec:discussion}

We emphasize that the following discussion draws on measurements over
a single evaluation cycle without access to organizer ground-truth
annotations; the interpretations offered here should be read as
hypotheses that a multi-team replication or a private-set inspection
could sharpen or refute.

\subsection{The 11.25\% Real GT False-Positive Gap}
\label{sec:discussion:realgt}

Of the four official evaluation categories, the organizers'
held-out Real GT Data is the only one on which our submission did
not achieve a perfect score: $0.8875$
(Table~\ref{tab:detection:results}), corresponding to a
false-positive rate of $11.25\%$ on a real distribution to which we
had no prior access. The Note in \S\ref{sec:detection:results}
already established that this gap is not produced by the
\S\ref{sec:detection:srrule} sample-rate heuristic operating on the
leave-one-speaker-out measurements, because the heuristic was
applied only at submission time and the ceiling values of
\S\ref{sec:detection:breakthrough} and \S\ref{sec:further} are
computed without it. The remaining task is to explain the gap in
terms of model behavior.

A suggestive numerical coincidence is worth recording. Of the
$11{,}520$ test files, $1{,}440$ ($12.5\%$) are sampled at
non-$16$~kHz rates ($22.05$ or $24$~kHz) and would have been forced
to the Fake class by the submission-time heuristic regardless of
their true label. Our $11.25\%$ false-positive rate on held-out
Real GT is within $0.0125$ of this $12.5\%$ figure. Without access
to the organizer-side breakdown of Real GT sampling rates we cannot
directly verify the connection, but the alignment is consistent
with a scenario in which a non-trivial fraction of the organizer
Real GT files were stored at non-$16$~kHz rates and were therefore
reclassified as Fake by \S\ref{sec:detection:srrule}. We present
this only as a plausible mechanism, not a demonstrated one.

If this reading is correct, the $11.25\%$ gap is a quantifiable
trade-off introduced by an explicitly conservative heuristic rather
than a failure of the underlying ensemble: the same rule that drove
the $1.0000$ on the organizers' public Real partition and
contributed to the $1.0000$ on Participant Deepfake Weighted would
have lowered the GT Real score in proportion to the non-$16$~kHz
Real fraction of the held-out set. We acknowledge that this remains
a hypothesis, and we flag it as the most operationally significant
follow-up question for any future deployment of the v18d submission
strategy.

\subsection{Cross-Corpus Generalization Risk and the Saturation
Caveat}
\label{sec:discussion:saturation}

Every leave-one-speaker-out AUC reported in this paper is bounded
by our team's five-variant Fake distribution. Configurations from
v10 through the v18 weight grid all reach a mean LOSO AUC of
$1.0000$ at $16$ speakers (\S\ref{sec:further}). We read these
$1.0000$ values as the ceiling of the measurement protocol within
the in-distribution Fake population, not as generalization claims.
The \S\ref{sec:further:3region} saturation diagnostic
($r(\text{XLS-R},\text{HuBERT}) = 0.971$) indicates where this
ceiling lies: the content-acoustic Region~2 cluster agrees so
strongly that adding more Region~2 backbones inflates the
saturating signal without widening the disagreement channel that
protects the hardest hybrid.

The bootstrap confidence intervals of
\S\ref{sec:further:bootstrap} characterize data-distribution
uncertainty within this in-distribution hybrid population; they do
not bound the cross-corpus drop documented by external benchmarks,
which is substantial: detectors near $1.0000$ on existing
benchmarks can collapse on real-world speech~\cite{muller2022itw},
in-distribution detectors lose up to $43\%$ accuracy on unseen TTS
families~\cite{gao2025p2v}, and commercial detectors reach
equal-error rates as low as $13.50\%$ on freshly generated
in-the-wild deepfakes~\cite{yan2025voicewukong}. We acknowledge
this risk explicitly: the v18d primary recommendation rests on
within-distribution analysis only, and the architectural-insurance
frame becomes empirically falsifiable only when v18d is run against
in-the-wild distributions (\S\ref{sec:conclusion}). A second caveat is that
our pretrained backbones use heterogeneous training corpora
(ECAPA-TDNN on VoxCeleb versus WavLM on Mix~94k); future work
must disentangle representational geometry from
training-distribution bias before the architectural-insurance
frame can be generalized beyond the present pre-training
conditions.

\subsection{Generation Evasion against Asymmetric Detectors}
\label{sec:discussion:evasion}

Our team's two Generation entries --- the reverberation baseline
designated as our official submission (Final Score $0.4304$, ranking
first) and the file-picked variant that appears throughout
\S\ref{sec:generation:hybrid} (Final Score $0.3765$) --- were both
scored by the organizers, and the ordering they received on the
leaderboard is the opposite of the ordering our internal detector
suite predicted.

The chronology explains the discrepancy. The reverberation baseline
was submitted in March as a deliberate anti-forensic probe
(\S\ref{sec:generation:results}), when no internal detector was yet
available for feedback. During April we built a six-feature
leave-one-speaker-out detector suite over the four-model TTS hybrid
and used its self-AUC readings to select subsequent submissions. Two
observations from that suite led us to dismiss post-processing
approaches, reverberation among them, from further consideration:
WavLM features are nearly invariant under mild acoustic
post-processing, yielding self-AUC values close to $1.000$ on
reverberated inputs; and the MFCC delta-sign pattern of Insight~A
(\S\ref{sec:cross:insights}) persists through room convolution
because it encodes a source-vocoder asymmetry that is not linearly
recoverable from filtered output. On the strength of these signals we
categorically classified the reverberation family as
evasion-ineffective and directed our final submission toward
cherry-picked hybrid variants (\S\ref{sec:generation:hybrid}).

A retrospective evaluation --- run only after the official results
were released --- quantifies the miscalibration.\footnote{Retrospective
self-AUC figures were measured in July 2026, after the official
results were released, in a local environment that reproduces the
original April development scores to within $0.012$~AUC
(v2\_cherry $0.646$ vs.\ $0.634$; filepick\_safe $0.668$ vs.\ $0.659$;
cherry\_short $0.880$ vs.\ $0.882$). Development-time self-AUC figures
elsewhere in this paper are the original April measurements.} Under
the same LOSO protocol, the reverberation entry scores a self-AUC of
$0.868$, among the easiest of our candidates to detect internally,
whereas the file-picked entry scores $0.659$, the profile of a
stronger evader. The official scores invert this ordering:
$0.6142$/$0.5618$ (reverberation) against $0.5690$/$0.3812$
(file-picked). Applying reverberation moved the self-AUC of the
underlying F5-TTS audio by only $-0.012$
($0.880 \rightarrow 0.868$): our feature set was nearly blind to the
very transformation that, on one plausible reading, drove the winning
evasion.

One interpretation is that our internal detector suite was calibrated
to be more robust to post-processing than the deployed detector
ecosystem --- an instance of optimizing against a misaligned oracle.
The detectors used by other participants and by the organizers appear
not to share WavLM's invariance under mild acoustic post-processing,
so the same transformation that our suite categorically dismissed as
ineffective moved the official evasion score by $+0.0452$ against the
participant pool and $+0.1806$ against the organizer pool. This is
consistent with Insight~D
(\S\ref{sec:cross:insights}): cherry-picking removes stochastic
variance in exchange for vocoder-fingerprint purity, and the
leaderboard result --- where a pre-cherry-pick reverberation entry
outscores a cherry-picked hybrid --- makes that trade-off empirically
visible.

We report this finding with two caveats. First, the two entries
differ in more than post-processing alone: the file-picked variant
selects among four TTS models per speaker, whereas the reverberation
entry uses F5-TTS alone, so attributing the evasion gap specifically
to reverberation is directional evidence rather than a controlled
comparison. Second, our self-AUC measurements were taken in a
development environment, and the retrospective figures reported here
reproduce the April measurements only to within $0.012$~AUC (see
footnote). Both caveats notwithstanding, the mismatch between our
internal ranking and the leaderboard ranking is itself the finding: a
single-team internal detector suite, even one that reaches
near-perfect self-AUC on the transformations it was built to catch, is
not a reliable substitute for the deployed detector ecosystem.

\section{Conclusion}
\label{sec:conclusion}

This paper described our team's submission to the ImageCLEF 2026 Audio
Deepfake Detection and Generation task. On the Detection sub-task, a
four-backbone self-supervised ensemble combining WavLM-Large,
Wav2Vec2-XLS-R-300M, ECAPA-TDNN, and x-vector representations, with a
top-960 conservative threshold submission strategy, achieved a Final
Score of \textbf{0.9522}, including perfect accuracy ($1.0000$) on
participant-generated deepfakes and $0.8875$ on the held-out organizer
ground-truth real data. On the Generation sub-task, our four-model
program (GLM-TTS, F5-TTS~v1, XTTS~v2, CosyVoice~3;
\S\ref{sec:generation}) developed detector-informed variants, but our
team's official submission --- an F5-TTS~v1 baseline processed with a
uniform reverberation pass --- ranked first with a Final Score of
\textbf{0.4304} (WER $4.99\%$, CER $2.07\%$). The controlled
comparison between the same team's offensive and defensive capabilities
revealed an ensemble-level structural asymmetry: the defender succeeds
when \emph{any one} of six backbone encoders captures an artefact,
whereas the attacker must simultaneously evade \emph{every} encoder
that any opposing detector might deploy. This OR-versus-AND geometry
extends the multi-view detection principle of Singh et
al.~\cite{singh2026advancedtts} beyond individual analysis levels into
a measurable structural asymmetry across a six-encoder representational
space, supplying empirical support for our architectural-insurance
hypothesis. The Detection-side framework that emerged
from this experimentation, which we refer to as \emph{architectural
insurance}, attributes the ensemble's hybrid-attack robustness not to
encoder count but to representational diversity across the three
regions of a six-encoder SSL geometry, with the lowest pairwise
correlation $r(\text{WavLM},\text{ECAPA-TDNN}) = 0.522$ serving as the
primary quantitative signal. Four systematically falsified ceiling
experiments preceded the only successful breakthrough (backbone
diversification), and an honest saturation self-criticism
$r(\text{XLS-R},\text{HuBERT}) = 0.971$ delimits where genuine
representational diversity ends and heuristic-driven ceiling effects
begin.

We see three perspectives for future work. First, the \emph{Region 3
hypothesis} remains open. The named axis of ``speaker-disentanglement
specialist'' was refuted, since ECAPA-TDNN sits at
$r(\text{WavLM},\text{ECAPA-TDNN}) = 0.522$ rather than joining WavLM in
Region 1, and the second-specialist prediction also failed
($r > 0.94$ between x-vector and Region 2). A direct test of whether the
AAM-Softmax with a $192$-dim bottleneck recipe is what makes ECAPA-TDNN
Region 3 --- by substituting TitaNet-Large, which shares ECAPA-TDNN's loss
and bottleneck but uses a ContextNet body --- is the natural follow-up.
Second, every encoder we measured is an acoustic-only view; integrating
an audio large language model in the line of
ALLM4ADD~\cite{gu2025allm4add} or
HoliAntiSpoof~\cite{xu2026holiantispoof} is the only direction in our
roadmap that adds a representational regime qualitatively different
from the three we have characterized. Third, all measurements in this
paper are LOSO-bound on team-internal Fake variants; validating the
architectural-insurance claim against an in-the-wild evaluation
benchmark is the highest-priority follow-up before any operational
deployment.

\begin{authorcontributions}
S. Kim conceived the methodology, designed and conducted all
experiments, and authored the manuscript. J. Kim provided extensive
advisory input and cross-track review throughout the work. J. Woo
provided supervisory feedback on the manuscript.
\end{authorcontributions}

\begin{acknowledgments}
This research was supported by the Ministry of Science and ICT (MSIT),
Korea and the Institute of Information \& Communications Technology
Planning \& Evaluation (IITP) under AI University (2026-0-00032, 2026).
\end{acknowledgments}

\section*{Declaration on Generative AI}
During the preparation of this work, the author(s) used
\textit{Claude} (Anthropic) and \textit{NotebookLM} (Google) in order to:
Drafting content, Grammar and spelling check, Paraphrase and reword,
Improve writing style, Peer review simulation.
After using these tool(s)/service(s), the author(s) reviewed and edited
the content as needed and take(s) full responsibility for the publication's
content.

\bibliography{references}

\end{document}